\documentclass[%
 reprint,
 amsmath,amssymb,
 aps,
prd,
]{revtex4-2}

\usepackage{graphicx}
\usepackage{dcolumn}
\usepackage{bm}
\usepackage{xcolor,graphicx,lipsum}

\begin{document}

\preprint{APS/123-QED}

\title{Spin period evolution and X-ray spectral characteristics of the SMC pulsar SXP 46.6}

\author{Aman Kaushik}
\email{aman.kaushik@tifr.res.in}
 \affiliation{Department of Astronomy and Astrophysics, Tata Institute of Fundamental Research, 1 Homi Bhabha Road, Colaba, Mumbai 400005, India} 
\author{Sayantan Bhattacharya}%
\email{sayantan.bhattacharya@tifr.res.in}
\affiliation{Department of Astronomy and Astrophysics, Tata Institute of Fundamental Research, 1 Homi Bhabha Road, Colaba, Mumbai 400005, India}%
\author{Sudip Bhattacharyya}
\email{sudip@tifr.res.in}
\affiliation{%
 Department of Astronomy and Astrophysics, Tata Institute of Fundamental Research, Colaba, Mumbai 400005, India 
}%
\affiliation{MIT Kavli Institute for Astrophysics and Space Research, 
Massachusetts Institute of Technology, Cambridge, MA, 02139, USA}
\date{\today}

\begin{abstract}
We characterize the Small Magellanic Cloud (SMC) pulsar SXP 46.6 using \textit{NuSTAR} observations conducted in 2017. The spin period ($P$) of this neutron star decreased from its discovered value of 46.6\,s to a value of 45.984(1)\,s, indicating a spin-up at the rate of $\dot{P} = -1.13\times10^{-9}$ s s$^{-1}$. This spin-up rate is used to calculate a high pulsar magnetic field value of $2.25\times10^{13}$~G.
This process also gives a low magnetic field value, which we rule out here by 
constraining the inner accretion disk radius to be less than the radius of the innermost stable circular orbit.
The pulse profile, analyzed in soft, hard, and broad X-ray bands, shows a double-peaked structure, consistent with pencil beam emission from two antipodal hot spots on the neutron star surface. We also perform spin phase-resolved spectroscopy for the first time, revealing spectral variations across different phases of the pulsar's rotation. These results offer new insights into the long-term spin evolution and emission properties of SXP~46.6.

\end{abstract}

\maketitle


\section{Introduction}\label{sec:intro}
X-ray binaries (XRBs) are astrophysical systems in which a compact object (a neutron star (NS) or black hole (BH)) accretes material from a companion or donor star, producing X-ray emission due to the gravitational energy released during the accretion process. 
XRBs are classified into two categories based on the companion star's mass: high-mass X-ray binaries (HMXBs) and low-mass X-ray binaries (LMXBs). 
HMXBs typically consist of a massive donor star, generally of O or B spectral type, with a compact object that is usually an NS, or in rarer cases, a BH \citep{1997xrb..book.....L}. 
In these systems, the donor star's strong stellar wind or decretion disk provides a reservoir of material for accretion onto the compact object, leading to persistent emission (e.g., SMC X-1 \citep{clarkson2003long}, etc.) or transient X-ray outbursts (e.g., SMC X-2 \citep{roy2022modeling}, X-3 \citep{koliopanos2018accreting}, etc.).

Be/X-ray binaries (BeXRBs) represent a major subclass of HMXBs where the companion star is a B-type star with a circumstellar decretion disk (see \citep{2011Ap&SS.332....1R} for a review on BeXRBs). These systems are characterized by regular X-ray outbursts linked to the interaction between the NS and the Be star's disk at periastron. The NS, typically in a wide and eccentric orbit, accretes material as it passes through the dense equatorial disk of the Be star, giving rise to Type I outbursts near periastron, as well as occasional and larger Type II outbursts that can occur independent of the location in orbital phase \citep{1986ApJ...308..669S, 2001A&A...377..161O}. The variability of the X-ray flux and the shape of the pulse profile can provide insights into the accretion geometry and the physical conditions of the accretion flow.

The Small Magellanic Cloud (SMC) is an ideal laboratory for studying BeXRBs due to its relative proximity and the large population of such systems \citep{2015MNRAS.452..969C, 2016A&A...586A..81H}. 
The SMC hosts a significantly higher number of BeXRBs per unit stellar mass compared to the Milky Way \citep{2010ApJ...716L.140A, 2009ApJ...697.1695A}, making it a key target for population studies and understanding the formation and evolution of these systems.
The pulsar SXP 46.6, originally discovered with the {\it Rossi X-ray Timing Explorer} ({\it RXTE}) in 1997 \citep{1997IAUC.6777....2M, 2008MNRAS.384..821M}, is one such BeXRB in the SMC. It exhibits characteristic X-ray pulsations and has shown a continuous decrease in its spin period over time, which is indicative of changes in the accretion torque. 
The position of the source was obtained from follow-up \textit{ASCA} observations, while \textit{RXTE} established a pulse period of 46.6\,s \citep{1998IAUC.6803....1C}. SXP~46.6 has been observed in the past decades with \textit{RXTE} and {\it Nuclear Spectroscopic Telescope Array} (\textit{NuSTAR}) with results presented in \citep{2008ApJS..177..189G, 2008MNRAS.384..821M, klus2014spin} showing regular outbursts helping in constraining the orbital period to 137.74 $\pm$ 0.42\,d.

In this paper, we present an analysis of SXP~46.6 based on \textit{NuSTAR} observations from 2017 \citep{2019ApJ...884....2L}, highlighting a significant spin-up from its initially discovered value of 46.6\,s to $45.984\pm{0.001}$\,s and comparison with previous such observations \citep{2008MNRAS.384..821M}. This spin period evolution provides an opportunity to study the torque exerted by the accretion flow onto the NS. 
Using the spin period derivative, we can also calculate the magnetic field of the NS \cite{1979ApJ...234..296G,klus2014spin}. Additionally, we perform spin period resolved spectroscopy for the first time and investigate phase-dependent spectral changes, analyze the pulse profile in soft, hard, and broad X-ray bands.

We describe the data and instrument in section~\ref{sec:obs_analysis}, and detail the spectral and timing analysis and results in section~\ref{sec:result}.
We discuss the implications of our results in section~\ref{sec:discussion} and conclude our work in section~\ref{sec:conclusion}.

\section{Observation and Data Analysis}\label{sec:obs_analysis}

\begin{figure*}
    \centering
    \includegraphics[width=\textwidth]{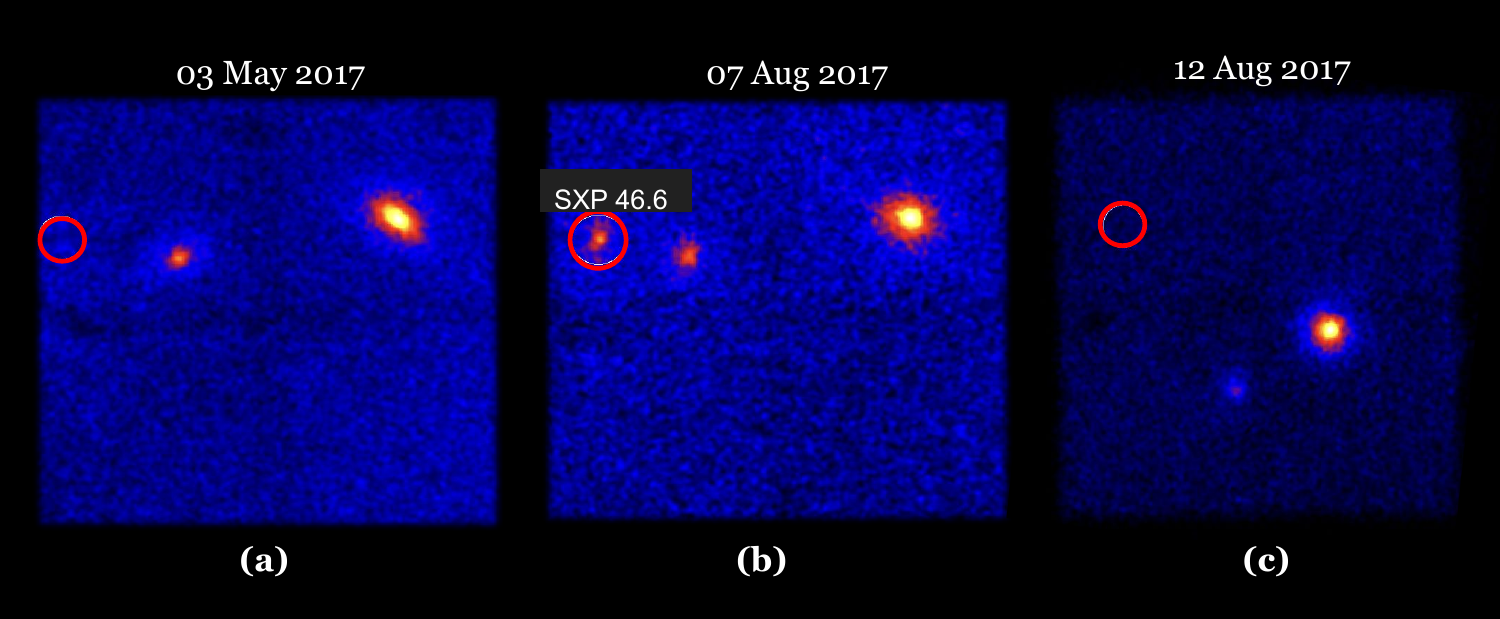}
\caption{This figure shows the evolution of the outburst of SXP~46.6 during its 2017 observation with \textit{NuSTAR} (see section~\ref{sec:obs_analysis}). It displays the field of view images obtained from the ObsIDs: (a) 50311003002, (b) 50311003004, and (c) 50311001004. 
The circular regions mark the position of SXP~46.6 in all the three panels.
The pulsar was undetected on $3^{rd}$ May 2017 and became visible on $7^{th}$ August 2017. The outburst faded and again became undetected in the observation after five days.}
    \label{fig:sxp_threepanel}
\end{figure*}

The \textit{NuSTAR} was launched on June 13, 2012, and is the first mission to use focusing telescopes to image the sky in hard X-rays (3--79 keV). 
The data for SXP~46.6 was collected in two separate $15' \times 15'$ fields. The source was in the field of view in three observations observed between 24 April 2017 and 12 August 2017 (see figure~\ref{fig:sxp_threepanel}--a, b, c).
The source was in the field of view but not detected for ObsID: 50311001004, 50311003002, and only got detected in ObsID: 50311003004. Hence, for this study, we have analyzed only ObsID: 50311003004, (see figure~\ref{fig:sxp_threepanel}b).


The data reduction for SXP~46.6 has been carried out using \textsc{heasoft} v6.34 and \textsc{caldb} v--20241126. \textit{NuSTAR} operates in the energy range of 3--79\,keV consisting of two identical X-ray telescope modules equipped
with independent mirror systems. Each of the telescopes has its focal plane modules A and B referred to as FPMA and FPMB respectively. For our spectral analysis we have only used data from FPMA in the energy range 3--20\,keV, as the counts were low and the spectrum had high error bars above 20\,keV. The data extraction and screening have been done using \textsc{nustardas}. The mission-specific \textsc{nupipeline} is run for filtering the unfiltered clean event files. Using \texttt{xselect v2.5b} and \texttt{ds9} application software, a circular region of $40''$ around the source center and that of the same size away from the source were selected as the source and background region files,  respectively. 
The selected source and background regions were utilized for extracting the necessary light curves and spectra by imposing the mission-specific \textsc{nuproducts}. 
For spectral fitting, we use the software \texttt{XSPEC v12.14.0h} and the $\chi^{2}$ statistics. We report the 1$\sigma$ uncertainty for the parameter values.

\section{Results \label{sec:result}}

\begin{figure}
    \centering
\includegraphics[width=\columnwidth]{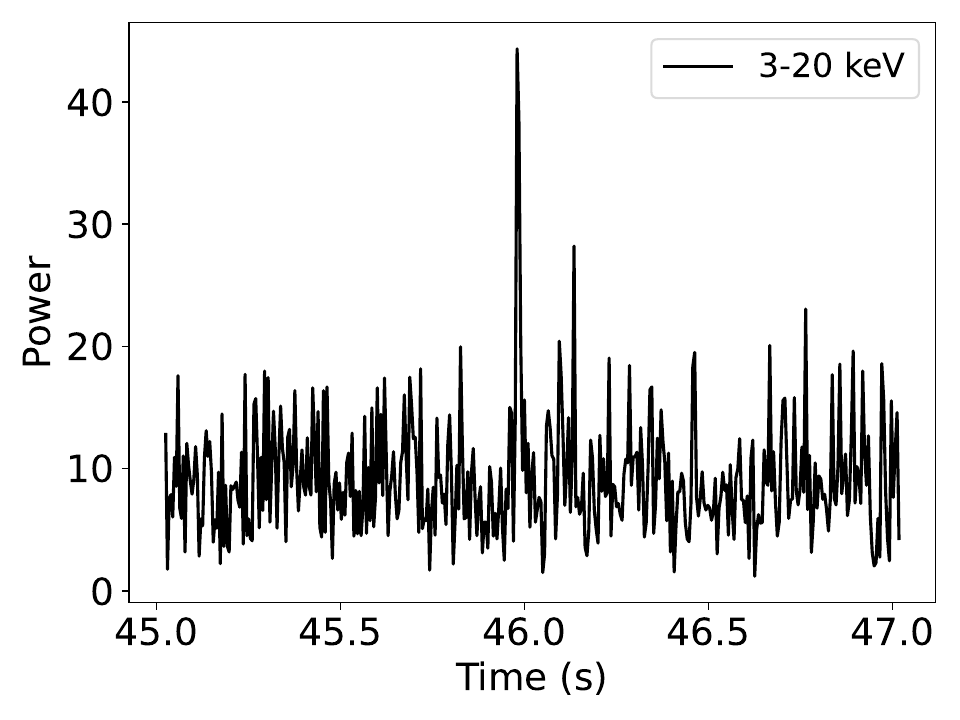}
\caption{The figure shows the spin period peak observed at $\sim$45.98\,s for the source SXP~46.6 observed in 2017 with \textit{NuSTAR} in 3--20\,keV energy range (section~\ref{sec:result}).}
    \label{fig:lomb_peak}
\end{figure}

\begin{figure}
    \centering
    \includegraphics[width=\columnwidth]{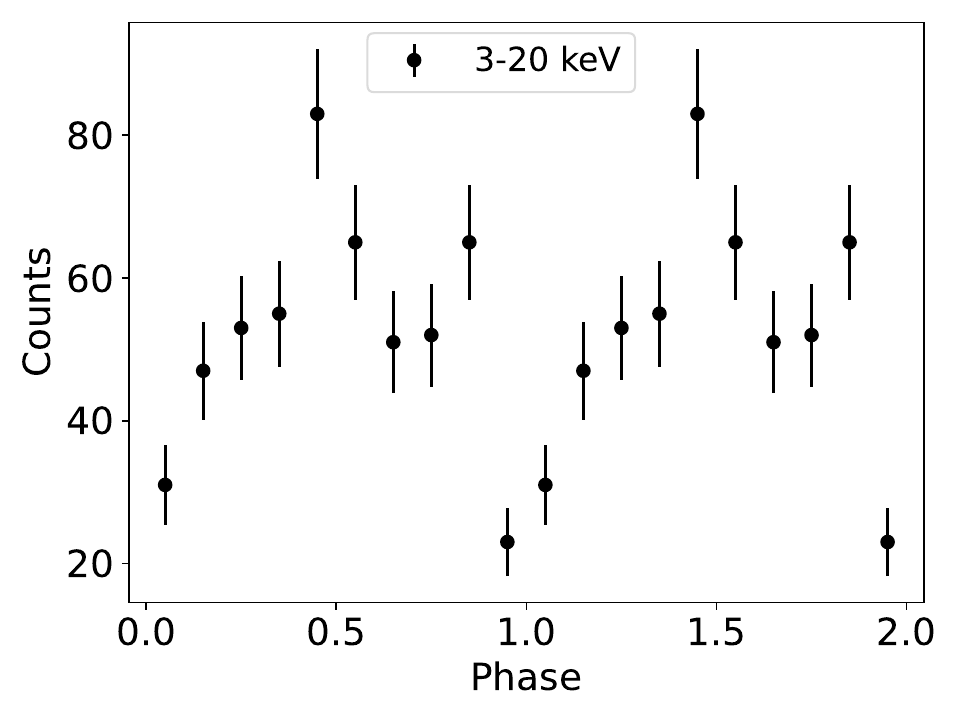}
    \includegraphics[width=\columnwidth]{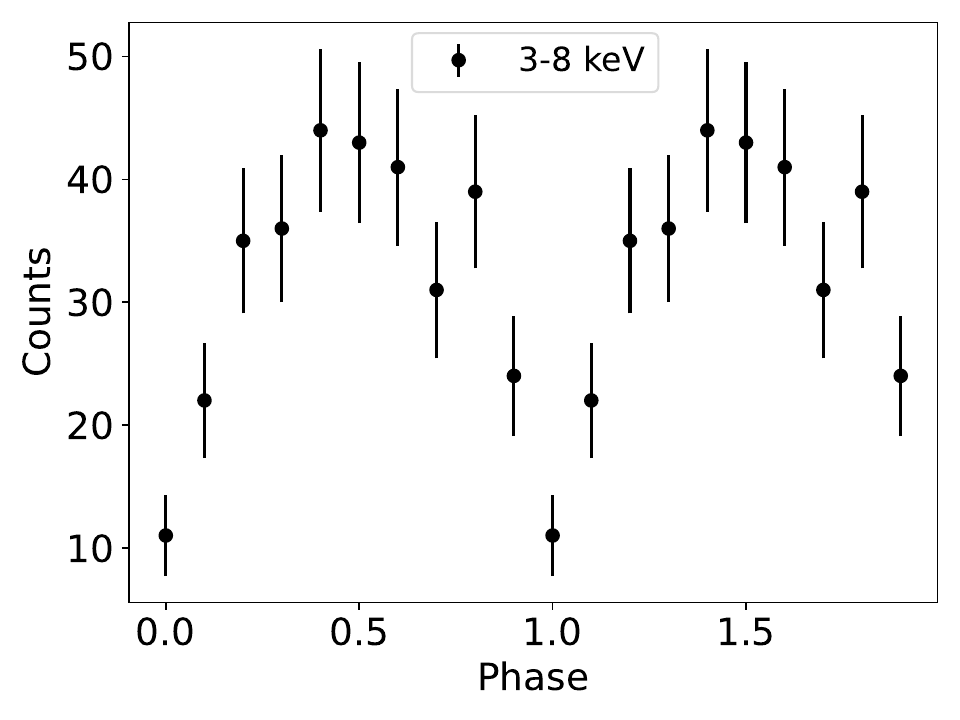}
    \includegraphics[width=\columnwidth]{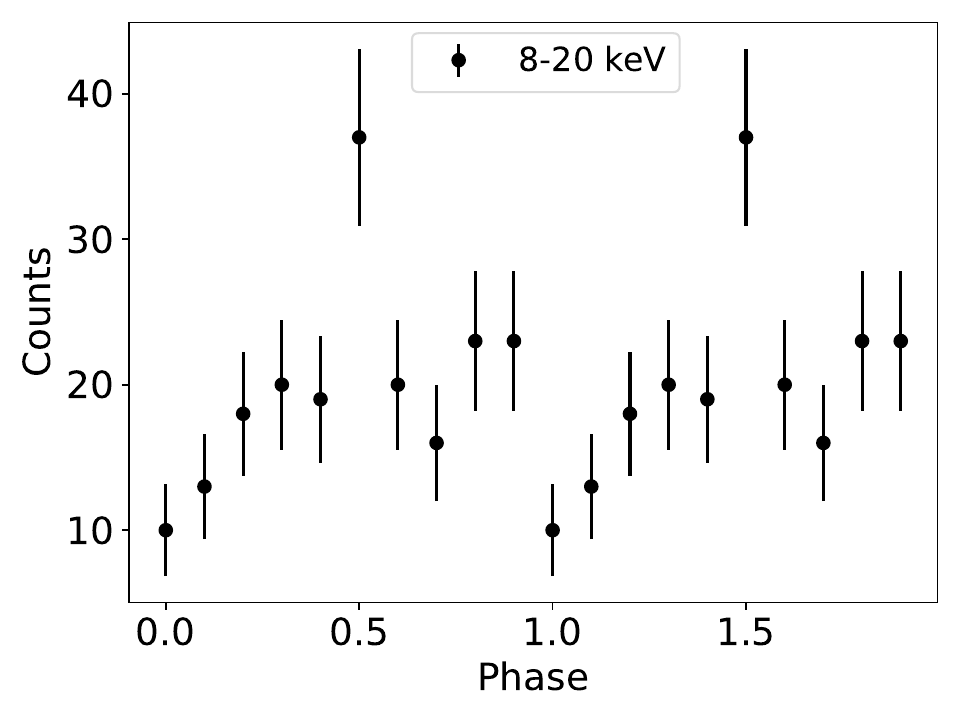}
\caption{The three panels in this figure show the pulse profile obtained by folding the light curve at $\sim$45.98\,s for three different energy ranges for the source SXP 46.6 observed with \textit{NuSTAR} during its 2017 outburst. The variation in the pulse profile hints towards a double peaked trend in all three panels }(section~\ref{subsec:duble-pulsed}).
    \label{fig:pulse_profile}
\end{figure}

To get a better understanding of the pulsations observed from this source, we analyze the temporal and phase-resolved spectral properties. To get an estimate of the pulsation period we use the \texttt{stingray.pulse.search} module from the software \texttt{STINGRAY v2.1}. 
\citep{Huppenkothen2019,2019ApJ...881...39H, matteo_bachetti_2024_11383212}. We calculate the uncertainty in the spin period value by using a bootstrapping approach combined with the above-mentioned periodogram analysis.
We filter the event file for an energy range of 3--20\,keV and search for the period in a frequency band centered on the previously known value of $\sim$46\,s, with a frequency bin size of the inverse of the observation time. 
We observe a significant peak around $45.984\pm{0.001}$\,s which is lower than the earlier observed value of $\sim$46.6\,s, hinting a spin-up of the pulsar due to the angular momentum transfer of the accreting material from the companion star to the NS  (Figure~\ref{fig:lomb_peak}).
Using these spin period values we estimate the rate of change in period as about $-0.0356$\,s\,$\mathrm{yr}^{-1}$. This is similar to the earlier calculated value \citep{2008MNRAS.384..821M}. 
We use this pulse period of $45.984\pm{0.001}$\,s to fold the light curve and generate pulse profiles having 10 bins in three energy ranges: 3--20\,keV, 3--8\,keV and 8--20\,keV (Figure~\ref{fig:pulse_profile}). We tried narrow energy bands: 3--6\,keV, 6--8\,keV, 8--12\,keV and 8--20\,keV but did not get enough SNR to generate a pulse profile. Hence,  we have chosen the current bands (3--20\,keV, 3--8\,keV and 8--20\,keV). The pulse profiles are
folded with a chosen time zero-point T$_0$ (folding epoch) so that the minimum flux bin lies at zero phase value. Figure~\ref{fig:pulse_profile} shows that the pulse profile in all three energy ranges hint towards a two peak trend which may indicate to two antipodal hot spots on the NS surface.

\begin{figure*}
    \centering
\includegraphics[width=\columnwidth]{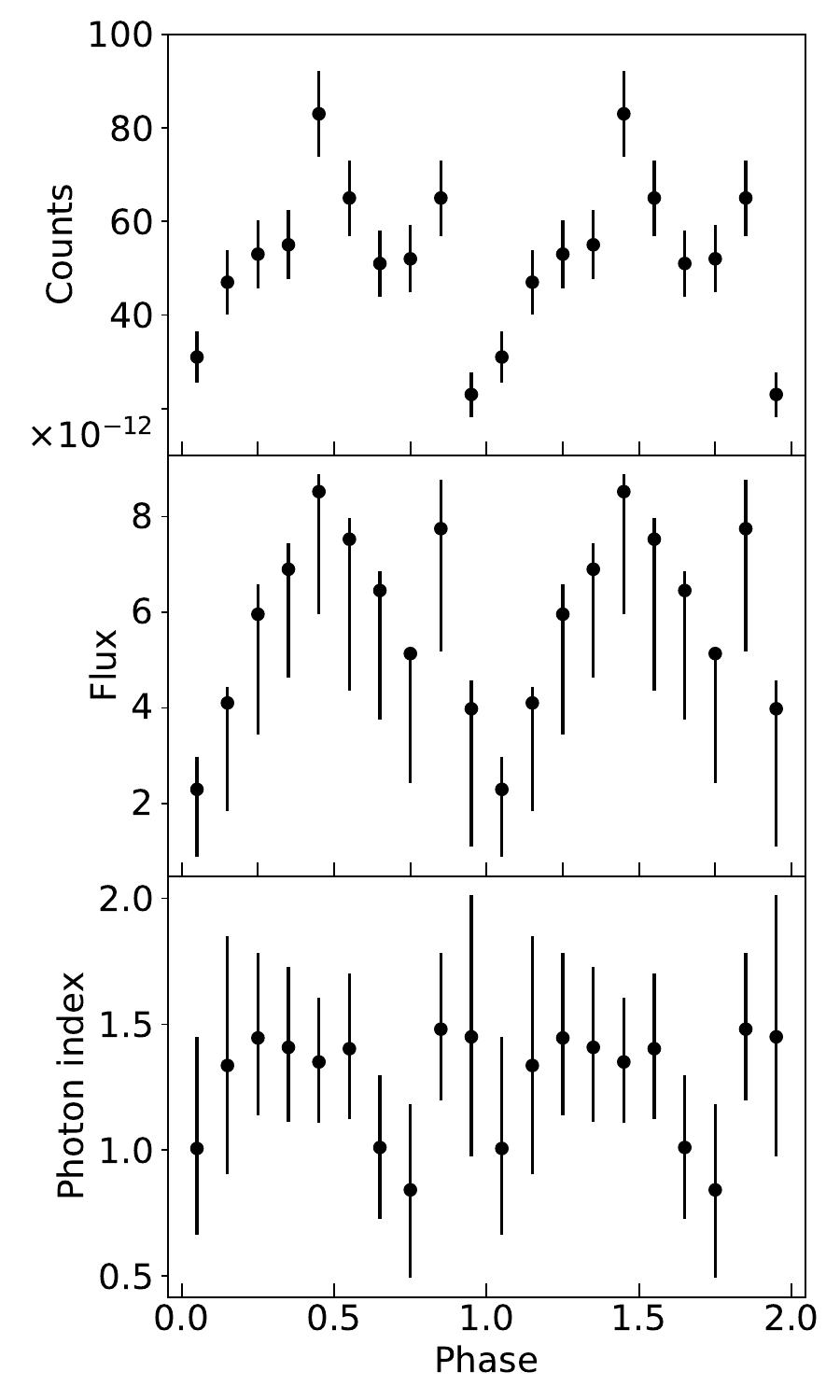}
\caption{This figure displays the variation of counts (top panel) in the energy range 3--20\,keV, flux (middle panel: because of the relatively low value of $n_{\mathrm{H}}$, both the unabsorbed and absorbed flux are similar) 
and photon index (bottom panel) with spin-phase in the energy range 3--20\,keV for the source SXP~46.6 observed with \textit{NuSTAR} during its 2017 outburst. The varying photon index indicates a positive correlation with the double-peaked pulse profile, (see section~\ref{subsec:phase-resolved}).}
    \label{fig:pulse_phase}
\end{figure*}

\begin{figure}
    \centering
\includegraphics[width=\columnwidth]{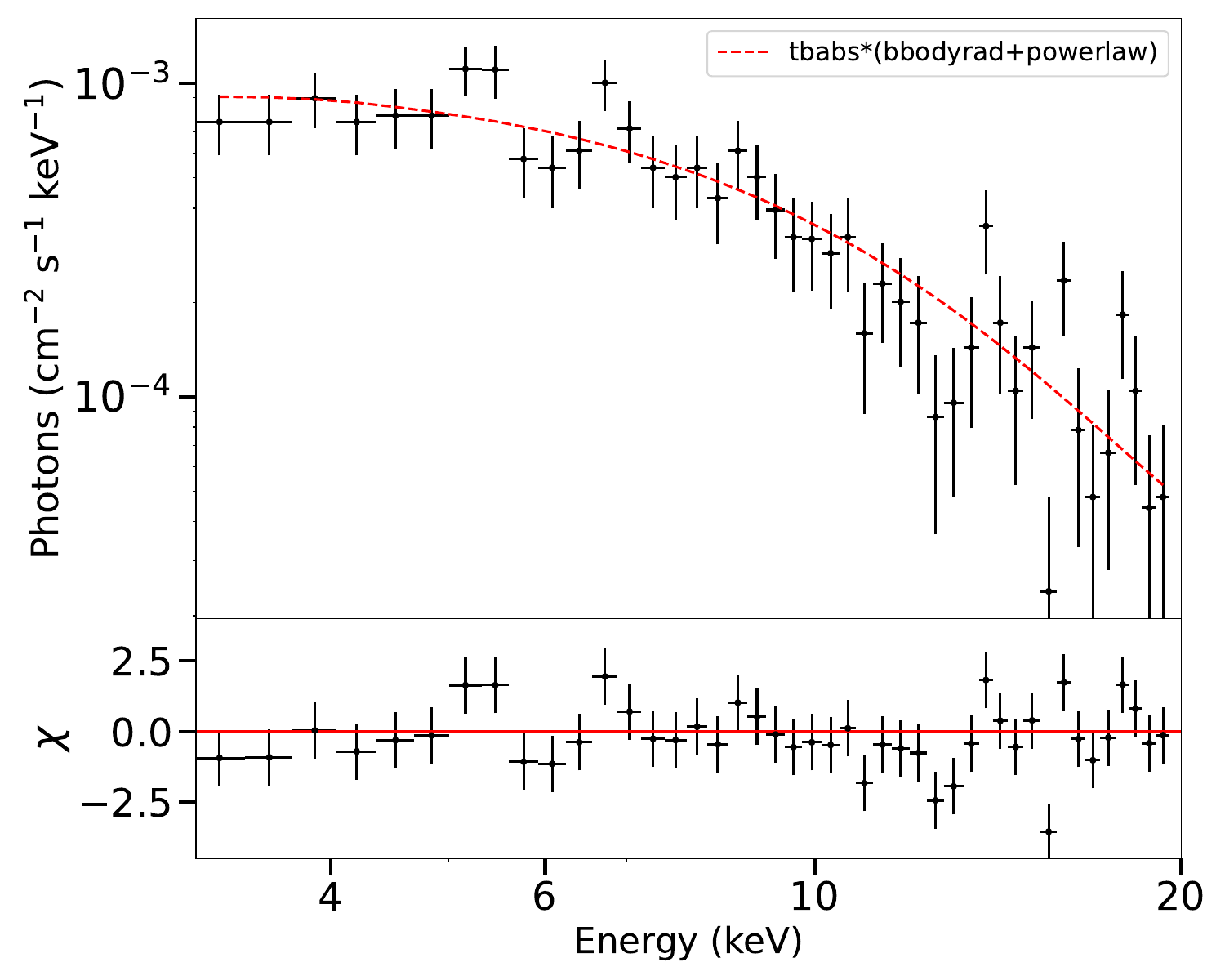}
 \caption{This figure shows the spectral fitting of the \textit{NuSTAR} data for the source SXP~46.6 during its 2017 outburst (ObsID: 50311003004). The spectrum is limited to 3--20\,keV energy range and is fitted using the \texttt{XSPEC} model \texttt{tbabs*(bbodyrad+powerlaw)} yielding $\chi^{2}$/dof = 45/41. The bottom panel shows the residual plot ($\chi$ = (Data - Model)/{error}); see section~\ref{sec:result}).}. 
\label{fig:spectral_fit}
\end{figure}

In order to analyze the anisotropic nature of the X-ray emission coming from the pulsar, we do phase-resolved spectral analysis.  A detailed spectral analysis has been done for the first time for this source.
For this, we divide the phase into 10 equal bins and filter the event file according to these bins using the software \texttt{XSPEC v12.14.0h}.
These phase-resolved event files are then used to extract the spectrum corresponding to each bin. Each spectrum is binned with the response file using the \texttt{ftgrouppha} (as part of \textsc{headas}) package. This method of binning is used to avoid over-sampling of counts as the spectrum is binned according to the energy resolution of the detector. The uneven binning across the spectrum is because of the grouping done using the algorithm, where at least a minimum number of counts are present in each bin. This algorithm creates broader bins, especially in harder energy ranges, to maintain the same counts throughout the spectrum.
Because of the low counts, each spectrum has high error bars above $\sim$20\,keV, hence we limit the spectrum in the energy range 3-20\,keV. From literature, we know that the emission in such systems is composed of thermal emission from the NS surface/accreting disk and non-thermal emission arising from the Compton up-scattering of thermal photons \citep{2022ApJ...939...67B}. Therefore, to describe the observed spectrum, we use the most simple, phenomenological models of thermal and non-thermal emission. We fit the spectra with the \texttt{XSPEC} model \texttt{tbabs}, to account for the absorption by interstellar and intrinsic neutral hydrogen medium and \texttt{powerlaw}, i.e., \texttt{tbabs*powerlaw}. The value of $n_\mathrm{H}$ is fixed to 0.49$\times10^{22}$\,cm$^{-2}$ during the fitting process, which is the average line of sight column density as measured from hydrogen intensity mapping \citep{2016A&A...594A.116H, 2005A&A...440..775K,1990ARA&A..28..215D}. Because of the low counts in each phase bin, adding another model component other than power-law doesn't affect the fit describing the spectrum.
We see a variation in the {\tt powerlaw} photon index value across the fitted spectra.
Figure~\ref{fig:pulse_phase} bottom panel shows this evolution with respect to the spin-phase, and the trend suggests a positive 
correlation with the pulse profile (top panel: counts; middle panel: flux), see section~\ref{subsec:phase-resolved}. 
The flux has been estimated in the 3–20\,keV energy range corresponding to each phase bin.
We also fit the full spectrum (all phases combined) of the source in the energy range 3--20\,keV (see figure~\ref{fig:spectral_fit}).
We start by fitting the simple model of power law (\texttt{tbabs*powerlaw}), but the fit is not satisfactory with a high reduced $\chi^{2}$ value ($\chi^{2}$/dof = 95/42).
This hints towards the requirement of a more complex model to describe the full spectrum.
The fit significantly improves by adding the \texttt{xspec} model \texttt{bbodyrad}, and the  \texttt{tbabs*(powerlaw+bbodyrad)} gives $\chi^{2}$/dof = 45/41. We calculate the 1$\sigma$ error in each model parameter using the \texttt{error} command in \texttt{XSPEC}. 
While calculating these 1$\sigma$ errors for the full spectrum fitting, the photon index becomes unconstrained and takes non-physical values. Therefore, we fix its value to the best fit result while calculating the errors of other parameter values (Table~\ref{tab:spectra_fit}). Additionally, since we use two different models to fit the full and phase-resolved spectra, the total flux values differ slightly.

\begin{figure}
    \centering
\includegraphics[width=\columnwidth]{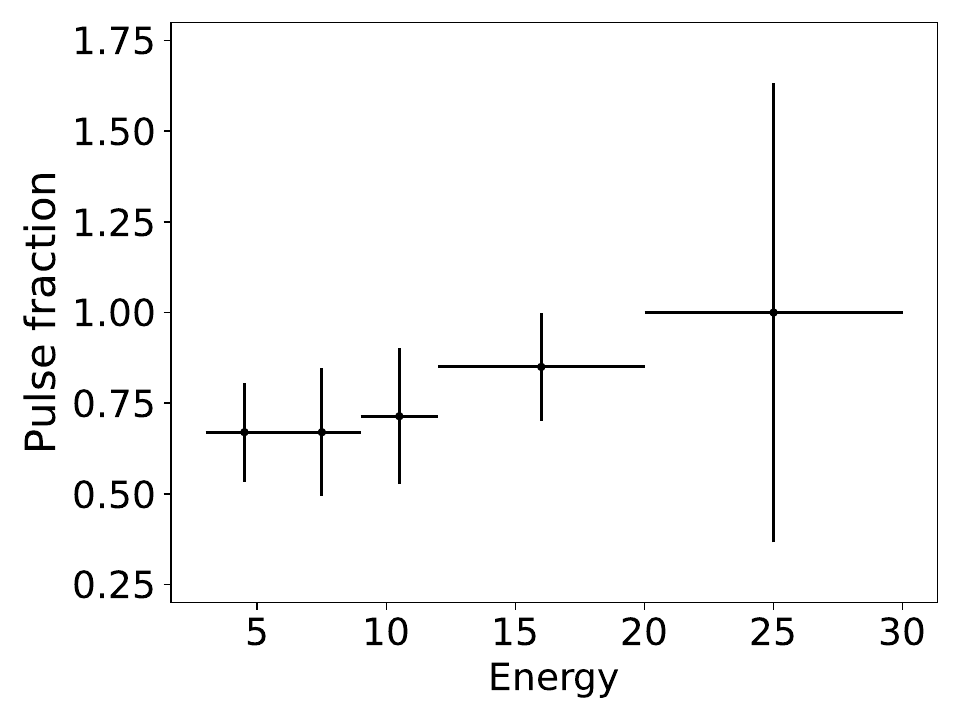}
\caption{This figure shows the variation of pulse-fraction with energy for the source SXP~46.6  during its 2017 outburst observed with \textit{NuSTAR} (see section~\ref{sec:result} for details).}
\label{fig:pulse_fraction}
\end{figure}

The pulse fraction (PF) represents the relative amplitude of the pulse profile. 
It relates X-ray emission from the accretion column (pulsed emission) and that from other regions of the accretion flow or NS surface (un-pulsed emission) \citep{2002ApJ...566L..85B}. It is defined as, PF = $(P_{\rm max} - P_{\rm min})/ (P_{\rm max}+P_{\rm min})$, where $P_{\rm max}$ and $P_{\rm min}$ represent maximum and minimum intensities of pulse profile, respectively. We calculate the pulse fraction in five energy ranges: 3--6\,keV, 6--9\,keV, 9--12\,keV, 12--20\,keV and 20--30\,keV. 
The PF plotted in Figure~\ref{fig:pulse_fraction} does not have a significant trend with energy.

\begin{table}
\centering
\caption{The best-fit spectral parameter values for the observations of SXP 46.6 during its 2017 outburst with \textit{NuSTAR} (section~\ref{sec:result}).}
\label{tab:spectra_fit}
\begin{tabular}{ccccc}
\hline
& powerlaw & bbodyrad & flux$^{\ddagger}$ $(10^{-12})$ & \\
\hline
\hline
Phase range & PhoIndex$^{\dagger}$ & $\text{k}T^{*}$\,(keV) & ${\mathrm{erg/cm^{2}/s}}$ & $\chi^{2}/\text{dof}$ \\
\hline 
All (0--1) & $1.41^{\text{fixed}}$ & $2.5\pm0.1$ & $67.2\pm0.1$ & 45/41 \\
0--0.1 & $1.01^{+0.44}_{-0.34}$ & -& $2.3^{+0.7}_{-1.4}$ & 31/34 \\
0.1--0.2 & $1.34^{+0.51}_{-0.43}$ & - & $4.1^{+0.3}_{-2.3}$ & 22/34 \\
0.2--0.3 & $1.45^{+0.34}_{-0.31}$ & - & $6.0^{+0.6}_{-2.5}$ &  18/34\\
 0.3--0.4 & $1.41^{+0.32}_{-0.29}$ & -& $6.9^{+0.5}_{-2.3}$ & 26/34 \\
0.4--0.5 & $1.35^{+0.25}_{-0.24}$ & -& $8.6^{+0.4}_{-2.5}$ & 22/35 \\
0.5--0.6 & $1.40^{+0.30}_{-0.30}$ & - & $7.6^{+0.4}_{-3.2}$ & 30/34\\
0.6--0.7 & $1.01^{+0.29}_{-0.28}$ & - & $6.5^{+0.4}_{-2.7}$ & 26/34\\
0.7--0.8 & $0.84^{+0.34}_{-0.35}$ & - & $5.2^{+0.1}_{-2.7}$ & 28/34\\
0.8--0.9 & $1.48^{+0.30}_{-0.28}$ & - & $7.8^{+1.0}_{-2.6}$ & 16/34\\
0.9--1.0 & $1.45^{+0.56}_{-0.48}$ & - & $4.0^{+0.6}_{-2.9}$ & 20/34\\
 \hline \hline
\end{tabular}
\begin{flushleft}
Notes: \\
The \texttt{XSPEC} model \texttt{tbabs*powerlaw} is applied to all the phase-resolved spectra, and the model \texttt{tbabs*(powerlaw+bbodyrad)} is applied to the full (phase range: 0--1) spectrum.
The value of $n_\mathrm{H}$ has been fixed to $4.9\times10^{21}\,\mathrm{cm}^{-2}$.\\
$\dagger$: The photon index of the powerlaw ({\tt powerlaw}) model.\\
$\ast$: The blackbody ({\tt bbodyrad}) temperature.\\
$\ddagger$: Since the value of $n_\mathrm{H}$ is relatively small, both the unabsorbed and absorbed flux have almost similar values.
\label{tab:time_lag}
\end{flushleft}
\end{table}

\section{Discussion}\label{sec:discussion}

In this paper, we present the analysis of \textit{NuSTAR} observations of the Small Magellanic Cloud (SMC) pulsar SXP~46.6. 
As shown in the previous section, we have significant findings that contribute to our understanding of its pulsed emission characteristics. 
The key results include the continuous spin-up of the pulsar leading to the reported spin period of $45.984\pm{0.001}$\,s, the detection of a double-peaked pulse profile indicative of antipodal hot spots with energy-dependent variations, and spin phase-resolved spectroscopy suggesting variation in the power-law index correlated with the pulse profile and exhibiting a phase lag. 
In the following subsections, we discuss the implications of these findings in the context of accretion-powered pulsars and NS magnetic field structure.

\subsection{Continuous Spin-Up of SXP~46.6}
The observed continuous spin-up of SXP~46.6 to a spin period of $45.984\pm{0.001}$\,s indicates an ongoing transfer of angular momentum from the accretion disk to the NS. Such behavior is commonly associated with high-mass X-ray binaries (HMXBs) in accretion regime, where the NS accretes matter from a companion star, in this case a Be-type star.

The continuous spin-up suggests a sustained mass accretion rate, possibly due to a decretion disk around the Be companion feeding material to the NS. This behavior aligns with the evolutionary patterns observed in other Be/X-ray binaries \citep{2011Ap&SS.332....1R}, where episodes of spin-up correspond to periods of enhanced accretion due to a well formed Be star's disk.


Furthermore, the observed spin period evolution also provides constraints on the NS's magnetic field strength. In order to get an estimate of this magnetic field of the NS in this system, we use the Ghosh and Lamb model \citep{1979ApJ...234..296G} of accretion on rotating magnetic NSs. We calculate the source luminosity by using the value of flux which is $\sim$ 7.74$\times10^{-11}$\,${\mathrm{erg/cm^{2}/s}}$ (3.0--20.0\,keV) and 60\,kpc as the distance to the source \citep{klus2014spin} and estimate it to be  $L \sim 3.3\times 10^{37}$\,erg s$^{-1}$. This corresponds to an accretion rate of $3.7\times10^{17}$\,g/s using the relation $L = \eta \dot{M} c^2$, taking $\eta=0.1$ \citep{2002apa..book.....F}.
According to this model, the rate of change in spin period ($P$) is given by the relation,

\begin{equation}
-\dot{P} = 5.0 \times 10^{-5} \mu_{30}^{2/7} n(\omega_s) R_{\text{NS}6}^{6/7} \left( \frac{M}{M_\odot} \right)^{-3/7} I_{45}^{-1} \left( P L_{37}^{3/7} \right)^2,
\end{equation}
where $\dot{P}$ is the rate of change of spin period measured in $\text{s}\,\mathrm{yr^{-1}}, \mu_{30} = \mu/10^{30}\,\mathrm{G}\,\mathrm{cm^{3}}, I_{45} = I/10^{45}\,\mathrm{g}\,\mathrm{cm^{2}}$, $R_{\text{NS}6} = R_{\text{NS}} /10^{6}\,\text{cm}$ and $L_{37} = L/10^{37}\,\mathrm{erg}\,\mathrm{s^{-1}}$. 
$R_{\text{NS}}$ is the radius of the NS, $I$ is the moment of inertia of NS and $\omega_s$ is the fastness parameter.
$n(\omega_s)$ is the dimensionless accretion torque and depends on the fastness parameter $\omega_s$ as
\begin{equation}
n(\omega_s) = 1.39 \left\{ 1 - \omega_s \left[ 4.03 (1 - \omega_s)^{0.173} - 0.878 \right] \right\} (1 - \omega_s)^{-1},
\end{equation}
for $0 < \omega_s < 0.9$, 
and, 
\begin{equation}
\omega_s = 1.35 \mu_{30}^{6/7} R_{\text{NS}6}^{-3/7} \left( \frac{M}{M_\odot} \right)^{-2/7} \left( P L_{37}^{3/7} \right)^{-1}.
\end{equation}
In these equations, we can see the degeneracy in $\omega_s$ values for a given $\dot{P}$ (see appendix~\ref{appendix}). 
For our source, for $\dot{P}$ $\sim -0.0356$\,$\text{s}\,\mathrm{yr^{-1}}$, and taking the NS mass as 1.4$M_\odot$ and radius as 12\,km, we get $10.8\times 10^{7}$\,G and $22.5\times 10^{12}$\,G as the NS magnetic field values corresponding to the two values of $\omega_s$. 

Previous studies \citep[e.g., ][]{klus2014spin} used the same formula and presented both low and high magnetic field values.
They could not exclude the lower values, which is not expected for HMXBs.
Here, we find a way to exclude the lower field value.
We show that for the lower field value the 
inner accretion
disk radius comes out to be less than the radius of the innermost stable circular orbit, which is not allowed (Figure~\ref{fig:risco_rkep}, appendix~\ref{appendix}).
Therefore, we estimate that the the NS magnetic field of the pulsar SXP 46.6 to be $2.25\times10^{13}$~G.

From this magnetic field value, we also estimate the centroid energy of the cyclotron line, which is also called cyclotron resonant scattering features (CRSFs) \citep{staubert2019cyclotron}. 
These are spectral features, that generally appear as absorption lines in the X-ray spectra of objects containing highly magnetic NSs. 
The value of the cyclotron line allows the direct measurement of the magnetic field strength in these objects. 
These features are thought to arise due to the resonant scattering of photons by electrons in the strong magnetic fields. 
To calculate the line value we used the relation

\begin{equation}
E_{cyc} = 11.6\left(\frac{B}{10^{12}}\right)\left(\frac{2GM}{Rc^{2}}\right)^{1/2}
\end{equation}
which for the above magnetic field value gives $E_{cyc} \approx 153.5\,\mathrm{keV}$. 
This value of $153.5\,\mathrm{keV}$ is outside the range of \textit{NuSTAR}, hence supports the non-detection here.


\subsection{Double-Peaked Pulse Profile and Antipodal Hotspots \label{subsec:duble-pulsed}}

The timing analysis results hint towards a double-peaked pulse profile in SXP~46.6, which may suggest the presence of two antipodal hotspots on the NS's surface, associated with the magnetic poles where accreted matter is funneled by strong magnetic fields. Although the peaks are not very significant in the energy range $>$8 keV. The pulse profile's shape ultimately provides insights into the geometry of the pulsar and the beam pattern of the X-ray emission \cite{cappallo2017geometry,cappallo2020geometry}.

We also see some evidence regarding one peak of the pulse profile being stronger in harder energies, the ratio of the two peaks increases for harder bands (but, based on one data point at phase = 0.5 of 8--20 keV pulse profile, see Fig.~\ref{fig:pulse_profile}). If found with more significance, this may imply an energy-dependent beaming or absorption effect. This can be attributed to several factors: The emission from each pole could have different beam patterns due to asymmetries in the magnetic field configuration or accretion region (hotspot) geometries, leading to variations in observed flux at different energies. Harder X-ray photons may originate from regions with higher optical depths or more significant bulk-motion comptonization, enhancing the hard X-ray emission from one pole over the other \cite{cappallo2020geometry}. Local absorption (i.e. CRSF feature) near one pole could attenuate soft X-rays more than hard X-rays in some cases, making the peak appear stronger at higher energies. However, no cyclotron lines are detected in this observation. 

The energy dependence of the pulse profiles provides valuable clues about the emission mechanisms and the magnetic field structure \cite{psaltis2014prospects}. By modeling the pulse profiles at different energies, constraints can be placed on the angles between the rotation axis, the magnetic axis, and the line of sight, as well as on the emission beam patterns (e.g., pencil vs.\ fan beams).

Comparisons with other pulsars exhibiting similar double-peaked profiles can shed light on whether this is a common feature among HMXBs \cite{laycock2024pulse} with certain properties or indicative of unique characteristics of SXP~46.6.

\subsection{Phase-Resolved Spectroscopy and Power-Law Index Variations \label{subsec:phase-resolved}}

 Initially, we fitted the phase-averaged (total) spectra where both power law and black body components are required to get a good fit, but the power law index needs to be frozen. Whereas this problem is resolved in phase resolved spectroscopy, where the flux variation is actually dictated by the power law and the black body component is not required (no change in reduced $\chi^2$ after adding blackbody component). We can infer that the bad fit in phase-averaged spectra is actually due to averaging over this power law index's phase variation rather than not having a $\sim$2.5 KeV blackbody (which was added to get a better fit). There are two reasons for this assumption, one of them being that the derived blackbody temperature is too high for an NS-HMXB system, and \textit{NuSTAR} being a hard energy telescope often provides a wrong estimate of this temperature unless jointly fitted with a soft X-ray telescope like \textit{NICER/Swift/XMM-Newton}. There might be a black body component too, it's just not well characterized with the present data. This motivated us to further explore the phase-resolved data. The phase-resolved spectroscopy suggests that the power-law photon index varies as a function of the pulse phase, and following a similar trend of variation as the pulse profile's double-peaked shape (see Fig.~\ref{fig:pulse_phase}). This variation may imply changes in the spectral hardness of the emitted X-rays associated with the rotation of the NS \cite{zhao2018pulse}.

The photon index $\Gamma$ is related to the emission processes in the accretion columns near the magnetic poles. Variations in $\Gamma$ may suggest that there can be changes in the physical conditions (e.g., temperature, density, magnetic field strength) within the accretion columns as they rotate into and out of the line of sight. The presence of two hotspots might contribute differently to the observed spectrum, with one producing harder spectra due to higher energy electrons or more efficient comptonization. A visual inspection of the Fig.~\ref{fig:pulse_phase} suggests that the photon index variation with phase lags behind the pulse profile. Although given the error bars in $\Gamma$, this claim cannot be strongly established. Still, we performed a cross-correlation analysis to quantify this lag, if any. This analysis was done by taking each of the peaks separately. We find a phase lag between the first peak of the pulse profile and $\Gamma$ profile to be $\sim0.03\pm0.01$ and zero lag in the second peak. This hints that $\Gamma$ variation possibly shows a phase lag with the flux variation (albeit larger uncertainties in $\Gamma$), also seen in other NS-HMXBs. These variations may indicate that the maximum spectral hardness does not coincide exactly with the maximum flux \cite{tobrej2025spectro, zhao2018pulse}. This could result from geometrical effects, such as the misalignment between the magnetic and spin axes, causing the hotspot emission to project differently at various rotation phases.

Understanding the phase-resolved spectral variations aids in constructing detailed models of the pulsar's emission geometry and the physical processes in the accretion columns.

\section{Conclusion}\label{sec:conclusion}
We have conducted timing and spectral analysis of a 2017 \textit{NuSTAR} observation of the BeXRB pulsar SXP 46.6, a system in the Small Magellanic Cloud. The following results provide us with a better understanding of the spin-accretion interaction in this system: 
\begin{itemize}    
\item The observed continuous spin-up indicates sustained accretion from the Be disk, which was used to determine the magnetic field of SXP 46.6 as \( 22.5 \times 10^{12} \) G for the higher value of the fastness parameter.
\item Based on the derived magnetic field, we predict a CRSF feature at 153.5 keV. The lower magnetic field value is excluded as it corresponds to the smaller fastness parameter. This lower $\omega_{s}$ value was ruled out by comparing the inner disk radius with \( R_{\text{isco}} \) (see appendix~\ref{appendix} for detail).
\item Indications of double-peaked pulse profile structure are observed across multiple energy bands, which may imply emission from two antipodal hot spots on the NS's surface. 
\item A possible variation of the power law index has been observed with spin phase using phase-resolved spectroscopy. The power law index might be correlated with the pulse profile, although the amplitude of their variation differs. This might suggest that not just the comptonization but also the geometry/area of different hotspots play a role in giving asymmetric height to the two peaks in the pulse profile.
    
\end{itemize}

\section{Acknowledgement}
We thank the anonymous referee for thoughtful comments which significantly improved this paper. Sudip Bhattacharyya acknowledges financial support by the Fulbright-Nehru Academic \& Professional Excellence Award (Research), sponsored by the U.S. Department of State and the United States-India Educational Foundation. We are thankful to the \textit{NuSTAR} team for efficient and continuous monitoring of the source. This research has made use
of \textit{NuSTAR}’s data obtained through the High Energy Astrophysics Science Archive Research Center (HEASARC) Online
Service, provided by the NASA/Goddard Space Flight Center. All the data analyzed for this study is available in the public archive in HEASARC.

\appendix 
\renewcommand{\thesection}{\Alph{section}}
\setcounter{figure}{0}
\renewcommand{\thefigure}{A\arabic{figure}}
\section{Physics of the fastness parameter ($\omega_{s}$)} \label{appendix}

\begin{figure*}
    \centering
    \includegraphics[width=\columnwidth]{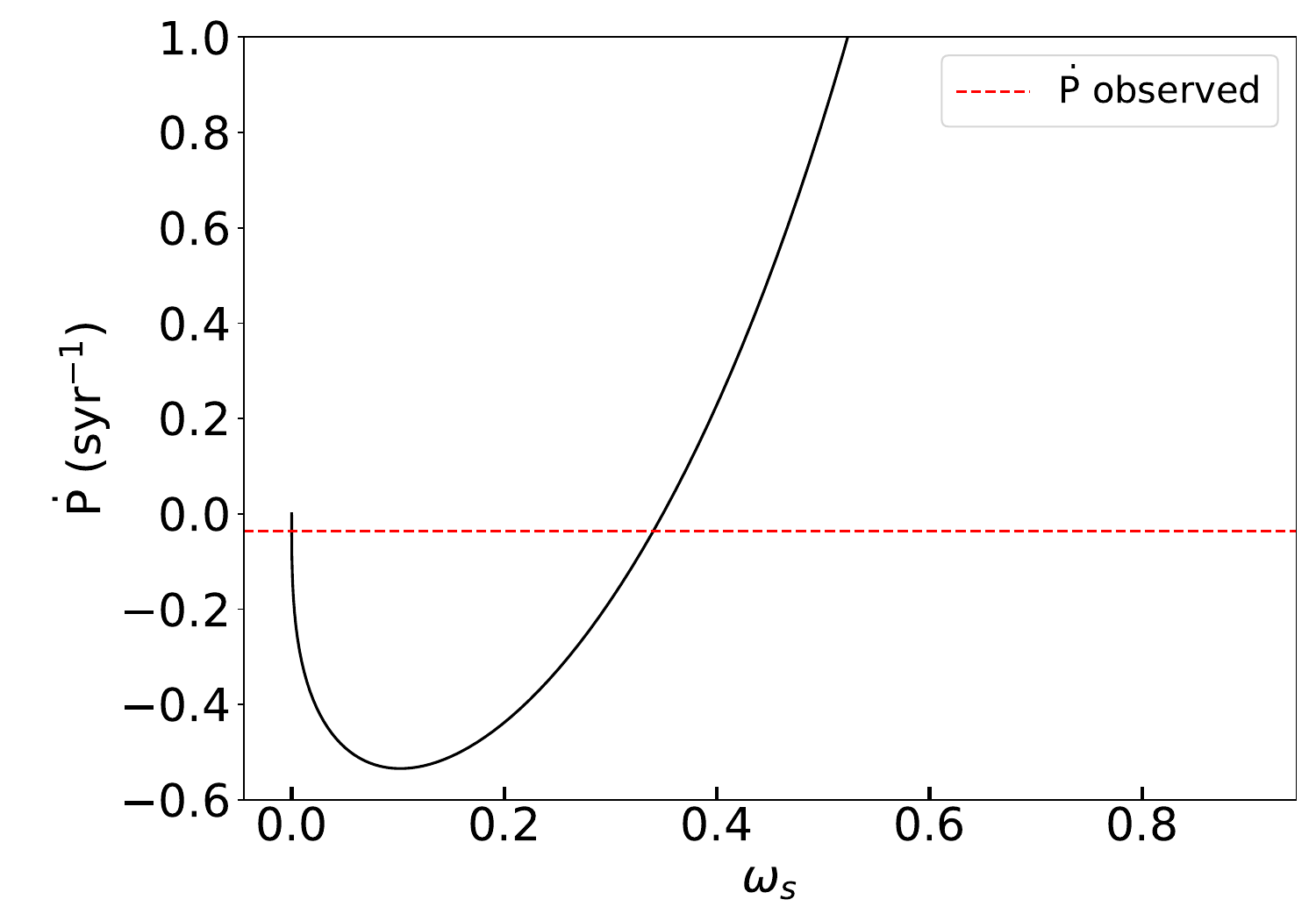}
    \includegraphics[width=\columnwidth]{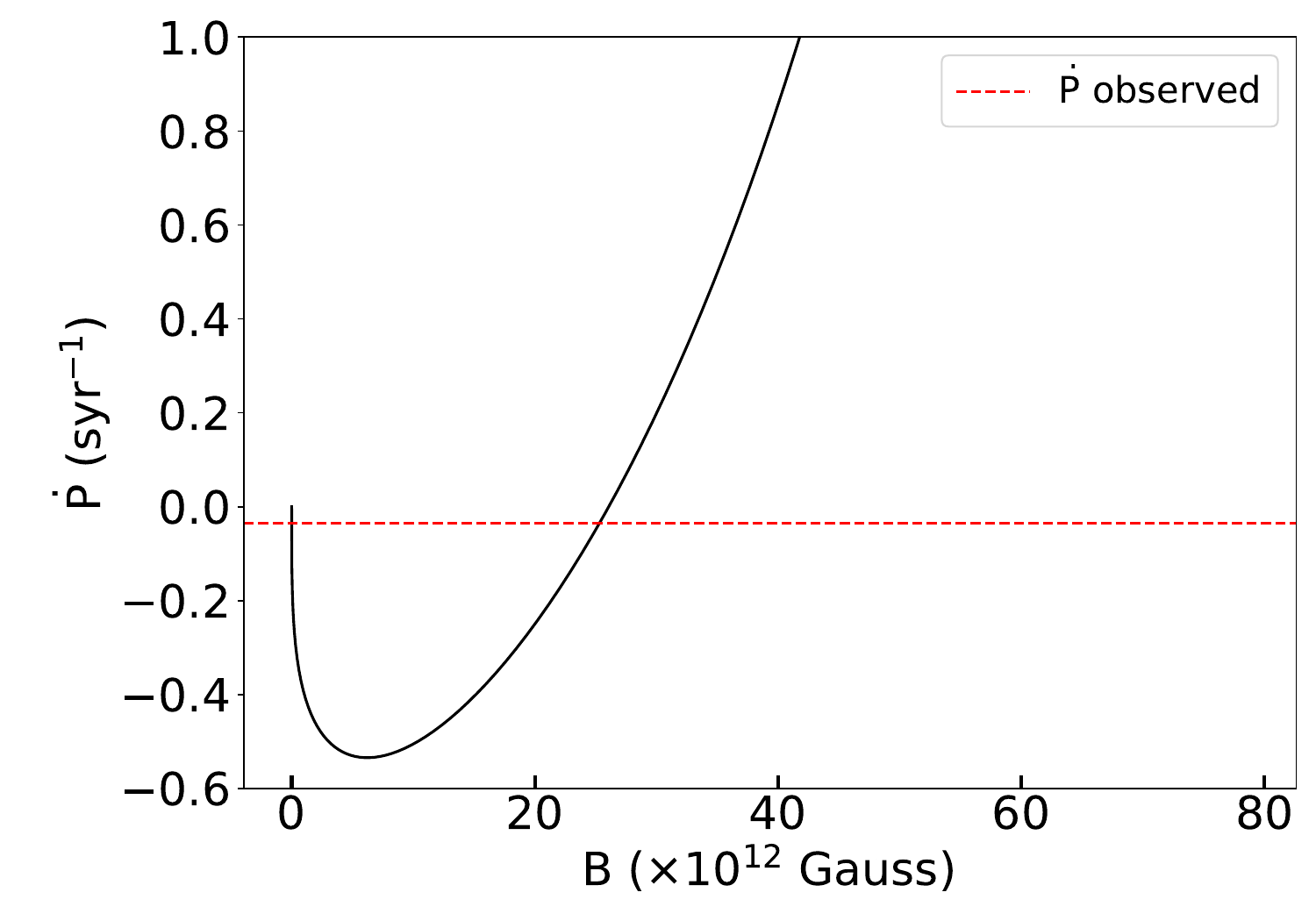}
    \caption{The above figure is for the source SXP 46.6 observed by \textit{NuSTAR} during its 2017 outburst. The plots display the evolution of fastness parameter $\omega_s$ with $\dot{P}$ (left panel) and magnetic field (B) with  $\dot{P}$ (right panel). Both the panels clearly show the degeneracy in both the parameters with respect to $\dot{P}$.}
    \label{fig:fast_para}
\end{figure*}

The fastness parameter ($\omega_{s}$) is a dimensionless quantity used to describe the interaction between an accreting NS's magnetic field and the accretion disk. It determines whether the NS is in the accretion or propeller regime. It is defined as:
\begin{equation}
\omega_s = \frac{\Omega_s}{\Omega_K(R_m)}
\end{equation}
where $\Omega_K(R_m)$ is the angular velocity of the accreting disk at the magnetospheric radius ($R_m$) and $\Omega_s$ is the angular velocity of NS \citep{1997MNRAS.286L..25L}. From Ghosh and Lamb model (1979) one can see the degeneracy between $\dot{P}$ and $\omega_s$ and 
$\dot{P}$ and magnetic field (B), see Figure~\ref{fig:fast_para}. The 2$^{nd}$ panel in the plot shows that for small values of $\omega_s$ we see degeneracy in magnetic field values, which means that for an observed $\dot{P}$ the model gives two values of NS magnetic field. One of those values is not physically allowed which can be confirmed by calculating the Keplerian radius of the accreting disk around the NS. For example for $\omega_s$ $\sim 10^{-6}$ the Keplerian radius is around 10\,km which is inside or touching the NS surface. For our analysis, we find two values of magnetic field corresponding to the $\dot{P}$ value of $\sim$ 0.0356\,s\,yr$^{-1}$, $\sim 1.1\times10^{8}\,\mathrm{G}$ and $22.5\times10^{12}\,\mathrm{G}$. The small B value is not physically feasible for the same above-stated reason and hence the larger value should reflect the true sense of the magnetic field around NS. Similarly, these low and high magnetic fields are also mentioned in \citep{klus2014spin} possibly coming from the same degenerate behavior. 

\begin{figure}
    \centering
    \includegraphics[width=\columnwidth]{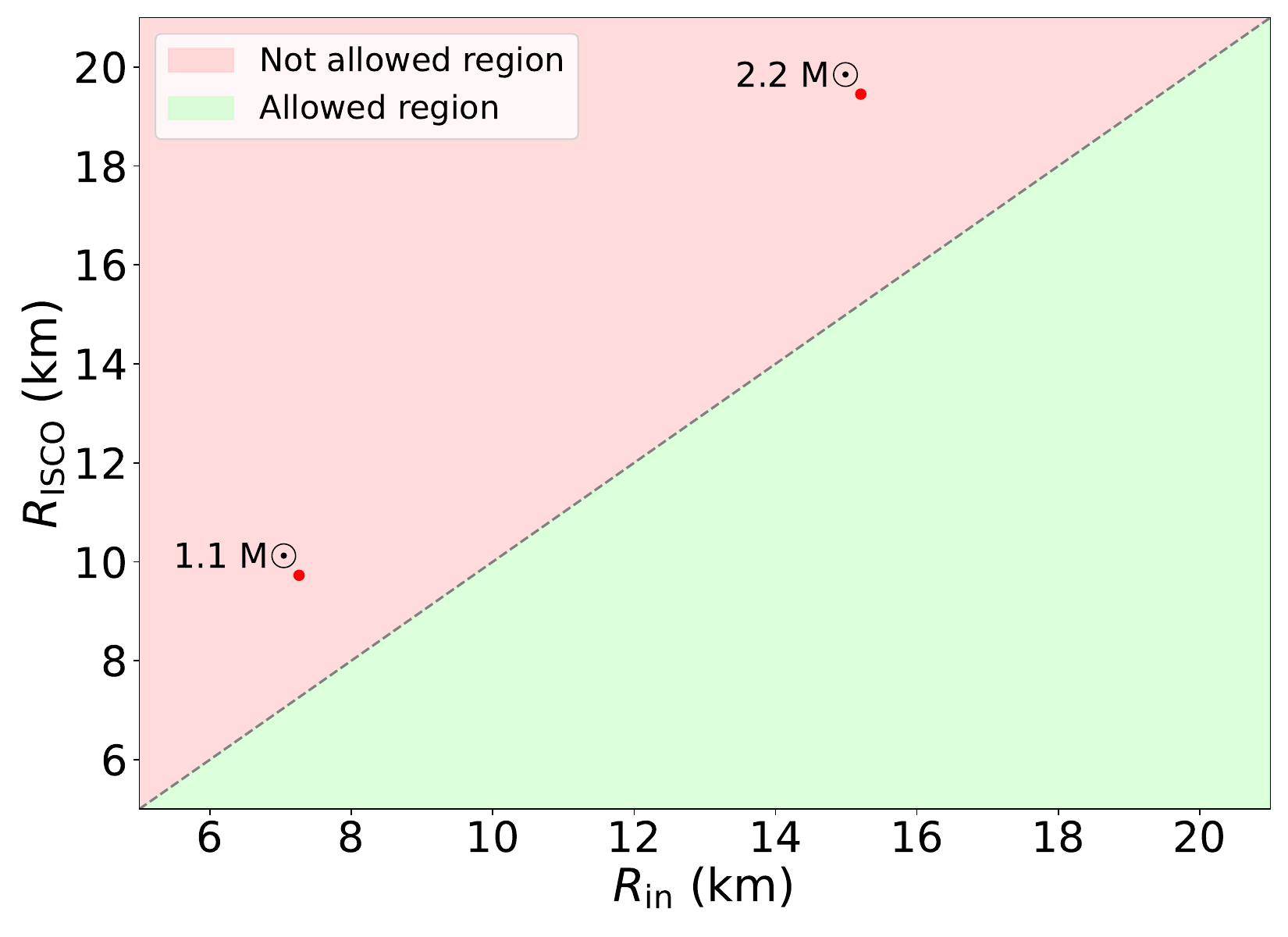}
    \caption{The above figure is for the source SXP 46.6 observed by \textit{NuSTAR} during its 2017 outburst. The figure shows the scatter plot between $R_\mathrm{ISCO}$ vs $R_\mathrm{in}$ (inner radius) for different masses of NS (only two have been shown) using the lower of the two degenerate values of $\omega_s$. We show that of the two degenerate values of $\omega_s$ for the observed $\dot{P}$ (-0.03563568\,$\text{s}\,\mathrm{yr^{-1}}$ in our case) the smaller $\omega_s$ and the corresponding values of magnetic field are not allowed. As for these $\omega_s$ values $R_\mathrm{ISCO}$ $>$ $R_\mathrm{in}$ (points lying above the straight dashed line and in the not allowed red shaded region). }
    \label{fig:risco_rkep}
\end{figure}

\nocite{*}
\bibliography{references}

\end{document}